\documentclass[11pt]{article}
\usepackage{graphics}
\usepackage{graphicx}
\usepackage{amsmath}
\usepackage{amsfonts}
\usepackage{amssymb}
\usepackage{amsthm}
\usepackage{tikz}
\usepackage{latexsym}
\usepackage{epstopdf}
\usepackage{amsbsy}
\usepackage{bbm}
\usepackage{color}
\usepackage[]{algorithm2e}
\usepackage{fullpage}
\usepackage{lineno}
\usepackage{enumitem}
\usepackage{subcaption}
\usepackage{comment}
\usepackage{hyperref}
\hypersetup{
    colorlinks = true,
    citecolor = blue,
    linkcolor = blue,
    urlcolor = black
}

\newtheorem{theorem}{Theorem}
\newtheorem{lemma}[theorem]{Lemma}

\newtheorem{prop}[theorem]{Proposition}
\newtheorem*{conj*}{Conjecture}

\theoremstyle{definition}

\begin{document}
%\linenumbers

\title{Ortho-unit polygons can be guarded with at most $\lfloor \frac{n-4}{8} \rfloor$ guards\footnote{The published version of this preprint can be found at \url{https://link.springer.com/article/10.1007/s00373-024-02880-8}, DOI: \url{https://doi.org/10.1007/s00373-024-02880-8}}}

\author{J.M. D\'iaz-B\'a\~nez 
\thanks{Departamento de Matem\'atica Aplicada II, Universidad de Sevilla, Spain. Supported in part by grants PID2020-114154RB-I00 and TED2021-129182B-I00 funded by MCIN/AEI/10.13039/50110001103
3 and the European Union NextGenerationEU/PRTR. {\tt dbanez@us.es} }
\and P. Horn\thanks{Department of Mathematics, University of Denver.  Supported in part by Simons Collaboration Grant 525039 {\tt Paul.Horn@du.edu}} \and M.A. Lopez %
\thanks{Department of Computer Science, University of Denver, {\tt mario.lopez@du.edu}. Supported in part by a University of Denver John Evans Research Award.}
\and N. Marín\thanks{Departamento de Ciencias de la Computaci\'on, Centro de Investigaci\'on Cient\'ifica y de Educaci\'on Superior de Ensenada, Baja California. {\tt nestaly@cicese.mx}. ORCID: 0000-0002-0222-3254. 
Supported by CONAHCYT of Mexico.}
\and A. Ramírez-Vigueras\thanks{Instituto de Matemáticas, UNAM. {\tt adriana.rv@im.unam.mx}. Partially supported by PAPIIT IN105221, Programa de Apoyo a la Investigación e Innovación Tecnologíca UNAM.}
\and O. Solé-Pi \thanks{Department of Mathematics, Massachusetts Institute of Technology.  \mbox{{\tt oriolsp@mit.edu}}.}
\and A. Stevens\thanks{Department of Computer Science, University of Denver. {\tt Alex.Stevens@du.edu}.} 
\and J. Urrutia \thanks{Instituto de Matem\'{a}ticas, Universidad Nacional Aut\'onoma de M\'{e}xico, {\tt urrutia@matem.unam.mx}. ORCID: 0000-0002-4158-5979. Supported in part by PAPIIT IN102117 Programa de Apoyo a la Investigaci\'on e Innovaci\'on Tecnol\'ogica, UNAM.}}

\maketitle

\begin{abstract}
An orthogonal polygon is called an ortho-unit polygon if its vertices have integer coordinates, and all of its edges have length one. In this paper we prove that any ortho-unit polygon with $n \geq 12$ vertices can be guarded with at most $\lfloor \frac{n-4}{8} \rfloor$ guards, which is a tight bound.
\end{abstract}

\section{Introduction}

A polygon $P$ is a sequence of $n$ vertices $v_1, \ldots , v_n$ and $n$ edges $v_1v_2, v_2v_3,\ldots , \allowbreak v_{n-1}v_n, v_nv_1$ such that no pair of consecutive edges are collinear, and no pair of non consecutive edges share a point. In this paper, and following standard notation in the Art Gallery literature, we will refer to $P$ as the region enclosed by the union of the vertices and edges of $P$; the boundary of $P$, denoted as $\partial P$, consists of the edges and vertices of $P$. Following O'Rourke's terminology~\cite{o1987art}, we will assume that $\partial P \subset P$. A point $p \in P$ guards a point $q \in P$ if the line segment $pq$ joining them is contained in $P$. 
A set $S$ of points in $P$ \emph{guards} $P$ if any point in $P$ is guarded by at least one point in $S$. A polygon is called an \emph{orthogonal polygon} if its edges are horizontal or vertical line segments.

In this paper, we will study guarding problems for \emph{ortho-unit polygons}, that is, orthogonal polygons all of whose edges have length one. See Figure~\ref{fig:example} for an example. Our main result is that $\lfloor \frac{n-4}{8} \rfloor$ points are always sufficient and sometimes necessary to guard any ortho-unit polygon. 

\begin{figure}[ht]
    \centering
    \begin{tikzpicture}[scale=.22, every node/.style={draw=none}]%[line cap=round,line join=round,>=triangle 45,x=1.0cm,y=1.0cm]
%\clip(-0.40858506988309334,-0.40285092046222243) rectangle (44.41973324371638,44.38068385836089);
\draw [] (4.,-20.)-- (6.,-20.);
\draw [] (6.,-20.)-- (6.,-18.);
\draw [] (6.,-18.)-- (8.,-18.);
\draw [] (8.,-18.)-- (8.,-16.);
\draw [] (8.,-16.)-- (6.,-16.);
\draw [] (6.,-16.)-- (6.,-14.);
\draw [] (6.,-14.)-- (4.,-14.);
\draw [] (4.,-14.)-- (4.,-12.);
\draw [] (4.,-12.)-- (6.,-12.);
\draw [] (6.,-12.)-- (6.,-10.);
\draw [] (6.,-10.)-- (8.,-10.);
\draw [] (8.,-10.)-- (8.,-8.);
\draw [] (8.,-8.)-- (10.,-8.);
\draw [] (10.,-8.)-- (10.,-10.);
\draw [] (10.,-10.)-- (12.,-10.);
\draw [] (12.,-10.)-- (12.,-12.);
\draw [] (12.,-12.)-- (14.,-12.);
\draw [] (14.,-12.)-- (14.,-10.);
\draw [] (14.,-10.)-- (16.,-10.);
\draw [] (16.,-10.)-- (16.,-8.);
\draw [] (16.,-8.)-- (18.,-8.);
\draw [] (18.,-8.)-- (18.,-6.);
\draw [] (18.,-6.)-- (20.,-6.);
\draw [] (20.,-6.)-- (20.,-4.);
\draw [] (20.,-4.)-- (18.,-4.);
\draw [] (4.,-20.)-- (4.,-18.);
\draw [] (4.,-18.)-- (2.,-18.);
\draw [] (2.,-18.)-- (2.,-16.);
\draw [] (2.,-16.)-- (0.,-16.);
\draw [] (0.,-16.)-- (0.,-14.);
\draw [] (0.,-14.)-- (-2.,-14.);
\draw [] (-2.,-14.)-- (-2.,-12.);
\draw [] (-2.,-12.)-- (0.,-12.);
\draw [] (0.,-12.)-- (0.,-10.);
\draw [] (0.,-10.)-- (2.,-10.);
\draw [] (2.,-10.)-- (2.,-8.);
\draw [] (2.,-8.)-- (0.,-8.);
\draw [] (0.,-8.)-- (0.,-6.);
\draw [] (0.,-6.)-- (-2.,-6.);
\draw [] (-2.,-6.)-- (-2.,-4.);
\draw [] (-2.,-4.)-- (-4.,-4.);
\draw [] (-4.,-4.)-- (-4.,-6.);
\draw [] (-4.,-6.)-- (-6.,-6.);
\draw [] (-6.,-6.)-- (-6.,-8.);
\draw [] (-6.,-8.)-- (-8.,-8.);
\draw [] (-8.,-8.)-- (-8.,-10.);
\draw [] (-8.,-10.)-- (-10.,-10.);
\draw [] (-10.,-10.)-- (-10.,-12.);
\draw [] (-10.,-12.)-- (-12.,-12.);
\draw [] (-12.,-12.)-- (-12.,-14.);
\draw [] (-12.,-14.)-- (-10.,-14.);
\draw [] (-10.,-14.)-- (-10.,-16.);
\draw [] (-10.,-16.)-- (-8.,-16.);
\draw [] (-8.,-16.)-- (-8.,-18.);
\draw [] (-8.,-18.)-- (-6.,-18.);
\draw [] (-6.,-18.)-- (-6.,-20.);
\draw [] (-6.,-20.)-- (-4.,-20.);
\draw [] (-4.,-20.)-- (-4.,-22.);
\draw [] (-4.,-22.)-- (-2.,-22.);
\draw [] (-2.,-22.)-- (-2.,-24.);
\draw [] (-2.,-24.)-- (0.,-24.);
\draw [] (0.,-24.)-- (0.,-26.);
\draw [] (0.,-26.)-- (2.,-26.);
\draw [] (2.,-26.)-- (2.,-24.);
\draw [] (2.,-24.)-- (4.,-24.);
\draw [] (4.,-24.)-- (4.,-22.);
\draw [] (4.,-22.)-- (6.,-22.);
\draw [] (6.,-22.)-- (6.,-24.);
\draw [] (6.,-24.)-- (8.,-24.);
\draw [] (8.,-24.)-- (8.,-26.);
\draw [] (8.,-26.)-- (10.,-26.);
\draw [] (10.,-26.)-- (10.,-24.);
\draw [] (10.,-24.)-- (12.,-24.);
\draw [] (12.,-24.)-- (12.,-22.);
\draw [] (12.,-22.)-- (14.,-22.);
\draw [] (14.,-22.)-- (14.,-20.);
\draw [] (14.,-20.)-- (16.,-20.);
\draw [] (16.,-20.)-- (16.,-18.);
\draw [] (16.,-18.)-- (18.,-18.);
\draw [] (18.,-18.)-- (18.,-16.);
\draw [] (18.,-16.)-- (20.,-16.);
\draw [] (20.,-16.)-- (20.,-14.);
\draw [] (20.,-14.)-- (22.,-14.);
\draw [] (22.,-14.)-- (22.,-12.);
\draw [] (22.,-12.)-- (24.,-12.);
\draw [] (18.,-4.)-- (18.,-2.);
\draw [] (18.,-2.)-- (16.,-2.);
\draw [] (16.,-2.)-- (16.,-4.);
\draw [] (16.,-4.)-- (14.,-4.);
\draw [] (14.,-4.)-- (14.,-6.);
\draw [] (14.,-6.)-- (12.,-6.);
\draw [] (12.,-6.)-- (12.,-4.);
\draw [] (12.,-4.)-- (10.,-4.);
\draw [] (10.,-4.)-- (10.,-2.);
\draw [] (10.,-2.)-- (8.,-2.);
\draw [] (8.,-2.)-- (8.,0.);
\draw [] (8.,0.)-- (10.,0.);
\draw [] (10.,0.)-- (10.,2.);
\draw [] (10.,2.)-- (12.,2.);
\draw [] (12.,4.)-- (10.,4.);
\draw [] (10.,4.)-- (10.,6.);
\draw [] (10.,6.)-- (8.,6.);
\draw [] (8.,6.)-- (8.,8.);
\draw [] (8.,8.)-- (6.,8.);
\draw [] (6.,8.)-- (6.,10.);
\draw [] (6.,10.)-- (4.,10.);
\draw [] (4.,10.)-- (4.,8.);
\draw [] (4.,8.)-- (2.,8.);
\draw [] (2.,8.)-- (2.,6.);
\draw [] (2.,6.)-- (4.,6.);
\draw [] (4.,6.)-- (4.,4.);
\draw [] (4.,4.)-- (6.,4.);
\draw [] (6.,4.)-- (6.,2.);
\draw [] (6.,2.)-- (4.,2.);
\draw [] (4.,2.)-- (4.,0.);
\draw [] (4.,0.)-- (2.,0.);
\draw [] (2.,0.)-- (2.,-2.);
\draw [] (2.,-2.)-- (0.,-2.);
\draw [] (0.,-2.)-- (0.,0.);
\draw [] (0.,0.)-- (-2.,0.);
\draw [] (-2.,0.)-- (-2.,2.);
\draw [] (-2.,2.)-- (-4.,2.);
\draw [] (-4.,2.)-- (-4.,0.);
\draw [] (-4.,0.)-- (-6.,0.);
\draw [] (-6.,0.)-- (-6.,-2.);
\draw [] (-6.,-2.)-- (-8.,-2.);
\draw [] (-8.,-2.)-- (-8.,0.);
\draw [] (-8.,0.)-- (-10.,0.);
\draw [] (-12.,2.)-- (-12.,4.);
\draw [] (-12.,4.)-- (-10.,4.);
\draw [] (-10.,4.)-- (-10.,6.);
\draw [] (-10.,6.)-- (-8.,6.);
\draw [] (-2.,14.)-- (0.,14.);
\draw [] (0.,14.)-- (0.,16.);
\draw [] (0.,16.)-- (2.,16.);
\draw [] (2.,16.)-- (2.,14.);
\draw [] (2.,14.)-- (4.,14.);
\draw [] (4.,14.)-- (4.,16.);
\draw [] (4.,16.)-- (6.,16.);
\draw [] (6.,16.)-- (6.,14.);
\draw [] (6.,14.)-- (8.,14.);
\draw [] (8.,14.)-- (8.,12.);
\draw [] (8.,12.)-- (10.,12.);
\draw [] (10.,12.)-- (10.,10.);
\draw [] (10.,10.)-- (12.,10.);
\draw [] (12.,10.)-- (12.,8.);
\draw [] (12.,8.)-- (14.,8.);
\draw [] (14.,8.)-- (14.,10.);
\draw [] (14.,10.)-- (16.,10.);
\draw [] (16.,10.)-- (16.,8.);
\draw [] (16.,8.)-- (18.,8.);
\draw [] (18.,8.)-- (18.,6.);
\draw [] (18.,6.)-- (20.,6.);
\draw [] (20.,6.)-- (20.,4.);
\draw [] (20.,4.)-- (22.,4.);
\draw [] (22.,4.)-- (22.,2.);
\draw [] (22.,2.)-- (24.,2.);
\draw [] (24.,2.)-- (24.,0.);
\draw [] (24.,0.)-- (22.,0.);
\draw [] (24.,-12.)-- (24.,-10.);
\draw [] (24.,-10.)-- (22.,-10.);
\draw [] (22.,-10.)-- (22.,-8.);
\draw [] (22.,-8.)-- (24.,-8.);
\draw [] (24.,-8.)-- (24.,-6.);
\draw [] (24.,-6.)-- (26.,-6.);
\draw [] (26.,-6.)-- (26.,-4.);
\draw [] (26.,-4.)-- (24.,-4.);
\draw [] (24.,-4.)-- (24.,-2.);
\draw [] (24.,-2.)-- (22.,-2.);
\draw [] (22.,-2.)-- (22.,0.);
\draw [] (-10.,0.)-- (-10.,-2.);
\draw [] (-10.,-2.)-- (-12.,-2.);
\draw [] (-12.,-2.)-- (-12.,-4.);
\draw [] (-14.,-6.)-- (-16.,-6.);
\draw [] (-16.,-6.)-- (-16.,-4.);
\draw [] (-16.,-4.)-- (-18.,-4.);
\draw [] (-18.,-4.)-- (-18.,-2.);
\draw [] (-18.,-2.)-- (-16.,-2.);
\draw [] (-16.,-2.)-- (-16.,0.);
\draw [] (-16.,0.)-- (-14.,0.);
\draw [] (-14.,0.)-- (-14.,2.);
\draw [] (-14.,2.)-- (-12.,2.);
\draw [] (-12.,-4.)-- (-10.,-4.);
\draw [] (-10.,-4.)-- (-10.,-6.);
\draw [] (-10.,-6.)-- (-12.,-6.);
\draw [] (-12.,-6.)-- (-12.,-8.);
\draw [] (-12.,-8.)-- (-14.,-8.);
\draw [] (-14.,-8.)-- (-14.,-6.);
\draw [] (12.,2.)-- (12.,0.);
\draw [] (12.,0.)-- (14.,0.);
\draw [] (14.,0.)-- (14.,2.);
\draw [] (14.,2.)-- (16.,2.);
\draw [] (16.,2.)-- (16.,4.);
\draw [] (16.,4.)-- (14.,4.);
\draw [] (14.,4.)-- (14.,6.);
\draw [] (14.,6.)-- (12.,6.);
\draw [] (12.,6.)-- (12.,4.);
\draw [] (-8.,6.)-- (-8.,4.);
\draw [] (-8.,4.)-- (-6.,4.);
\draw [] (-6.,4.)-- (-6.,6.);
\draw [] (-6.,6.)-- (-4.,6.);
\draw [] (-4.,6.)-- (-4.,4.);
\draw [] (-4.,4.)-- (-2.,4.);
\draw [] (-2.,4.)-- (-2.,6.);
\draw [] (-2.,6.)-- (0.,6.);
\draw [] (0.,6.)-- (0.,8.);
\draw [] (0.,8.)-- (-2.,8.);
\draw [] (-2.,8.)-- (-2.,10.);
\draw [] (-2.,10.)-- (-4.,10.);
\draw [] (-4.,10.)-- (-4.,8.);
\draw [] (-4.,8.)-- (-6.,8.);
\draw [] (-6.,8.)-- (-6.,10.);
\draw [] (-6.,10.)-- (-8.,10.);
\draw [] (-8.,10.)-- (-8.,12.);
\draw [] (-8.,12.)-- (-6.,12.);
\draw [] (-6.,12.)-- (-6.,14.);
\draw [] (-6.,14.)-- (-4.,14.);
\draw [] (-4.,14.)-- (-4.,16.);
\draw [] (-4.,16.)-- (-2.,16.);
\draw [] (-2.,16.)-- (-2.,14.);
\end{tikzpicture}
    \caption{An ortho-unit polygon with $220$ vertices.}
    \label{fig:example}
\end{figure}

\section{Previous work}

The study of \emph{guarding} or illumination problems of polygons on the plane is a classic area of research in computational geometry. 
In 1973 V. Klee posed the problem of determining the minimum number of guards needed to guard any art gallery whose boundary is represented by a polygon (Honsberger, 1976~\cite{honsberger1976mathematical}). 
Klee's problem was solved by Chv\'atal~\cite{chvatal1975combinatorial}, who proved that $\lfloor \frac{n}{3} \rfloor$ guards are always sufficient and sometimes necessary. 
A short and elegant proof was later given by Fisk~\cite{fisk1978short} using triangulations of polygons. 

\begin{figure}[ht]%\label{Low-bounds}
%\captionsetup{width=.9\linewidth}
\centering

\begin{subfigure}[b]{0.4\textwidth}
\centering

\begin{tikzpicture}[scale=.45, every node/.style={draw=none}]
\draw[] (0,-1)--(10,-1);
\draw[] (0,-1)--(.5,2);
\draw[] (.5,2)--(1,0);
\draw[] (1,0)--(2,0);
\draw[] (2.5,2)--(2,0);
\draw[] (2.5,2)--(3,0);
\draw[] (3,0)--(4,0);

\node (a) at (5,.25) {$\cdots$};

\draw[] (6,0)--(7,0);
\draw[] (7.5,2)--(7,0);
\draw[] (7.5,2)--(8,0);
\draw[] (8,0)--(9,0);
\draw[] (9.5,2)--(9,0);
\draw[] (9.5,2)--(10,-1);

\end{tikzpicture}

\caption{}
\end{subfigure}
~
\begin{subfigure}[b]{0.4\textwidth}
\centering

\begin{tikzpicture}[scale=.45, every node/.style={draw=none}]
\draw[] (0,-1)--(0,0);
\draw[] (0,-1)--(10,-1);
\draw[] (10,-1)--(10,0);

\draw[] (0,0)--(0,1);
\draw[] (0,1)--(1,1);
\draw[] (1,0)--(1,1);
\draw[] (1,0)--(2,0);
\draw[] (2,1)--(2,0);
\draw[] (2,1)--(3,1);
\draw[] (3,0)--(3,1);
\draw[] (3,0)--(4,0);

\node (a) at (5,.25) {$\cdots$};

\draw[] (6,0)--(7,0);
\draw[] (7,1)--(7,0);
\draw[] (7,1)--(8,1);
\draw[] (8,0)--(8,1);
\draw[] (8,0)--(9,0);
\draw[] (9,1)--(9,0);
\draw[] (9,1)--(10,1);
\draw[] (10,0)--(10,1);
\end{tikzpicture}

\caption{}
\end{subfigure}

\caption{(a) is a family of polygons in which $\lfloor\frac{n}{3}\rfloor$ guards are required. (b) is a family of orthogonal polygons in which $\lfloor\frac{n}{4}\rfloor$ guards are required.}
\label{Low-bounds}
\end{figure}

For orthogonal polygons, Kahn, Klawa and Kleitman~\cite{kahn1983traditional} proved that $\lfloor \frac{n}{4} \rfloor$ points are always sufficient to guard any orthogonal polygon with $n$ vertices. This bound is also tight.
Since then, a plethora of papers studying several variations to Klee's original question have appeared; see~\cite{o1987art,shermer1992recent,urrutia2000art}.

Orthogonal polygons have been widely studied, among other reasons, since they model traditional buildings in a more realistic way. 
To our knowledge there are at least six different proofs of the Orthogonal Art Gallery Theorem, e.g. see~\cite{gyori1986short,o1983alternate,toussaint2014computational,urrutia1997sixth}. 
Hoffman and Kriegel~\cite{hoffmann1996graph} proved that any orthogonal polygon with holes can always be guarded with $\lfloor \frac{n}{3} \rfloor$ \emph{vertex guards}, i.e. guards placed at vertices of the polygon.
Bjorling-Sachs proved that $\lfloor \frac{3n+4}{16}\rfloor$ edge guards are always sufficient and sometimes necessary to guard an orthogonal polygon~\cite{bjorling1998edge}.
Katz and Roisman~\cite{katz2008guarding} proved that the problem of finding the minimum number of guards needed to guard the vertices of an orthogonal polygon is $NP$-hard.
Tom\'as~\cite{tomas2013guarding} proved that this problem is $NP$-hard even for a restricted family of polygons called \emph{thin} orthogonal polygons, orthogonal polygons such that their dual graph is a tree (the graph that joins adjacent regions of the partition obtained by extending the edges to the interior of the polygon until they hit its boundary).

The guarding problem for polyominoes, orthogonal polygons that are obtained by joining sets of unit squares, was studied by 
Biedl et al.~\cite{biedl2012art}. 
They proved that $\lfloor \frac{m+1}{3} \rfloor$ point guards are always sufficient and sometimes necessary to guard a polyomino with area $m$, possibly with holes. 
Katz and Morgenstern~\cite{katz2011guarding} studied the problem of guarding an orthogonal art gallery with cameras that slide along horizontal or vertical straight tracks. 
They proved that if only horizontal tracks are allowed, then finding the minimum number of tracks needed can be found in polynomial time. 
Durocher et al.~\cite{durocher2017guarding} extended Katz and Morgenstern results by proving that finding a set of tracks that minimizes the sum of their lengths can also be solved in polynomial time, even for orthogonal polygons with holes.
Biedl et al.~\cite{biedl2019guarding}
studied the problem of guarding orthogonal polygons with sliding $k$-transmitters, point guards that slide along an axis-parallel line segment $s$. 
A $k$ transmitter can see a point $p \in P$ if the line segment perpendicular to $s$ joining it to $p$ intersects the boundary of $P$ at most $k$ times. 
The interested reader can find more information on guarding problems in a book by O'Rourke~\cite{o1987art} and two survey papers, by Shermer~\cite{shermer1992recent} and Urrutia~\cite{urrutia2000art}.

\section{Definitions}

A \emph{vertical (resp. horizontal) cut} of an ortho-unit polygon $P$ is a vertical (resp. horizontal) segment joining two reflex vertices of $P$ whose interior lies completely in the interior of $P$.
As observed by O'Rourke in \cite{o1987art}, there are $\frac{n-4}{2}$ reflex vertices in any orthogonal polygon and hence there are $\frac{n}{4}-1$ vertical cuts.  
They, in turn, partition $P$ into $\frac{n}{4}$ one-wide rectangles, which we call \emph{columns}.  
Note that the top and bottom of each column are horizontal unit-length edges.  

\begin{figure}[!htb]
%\captionsetup{width=.8\linewidth}
\centering
\begin{subfigure}{0.65\textwidth}
\centering
\includegraphics[width=0.7\linewidth]{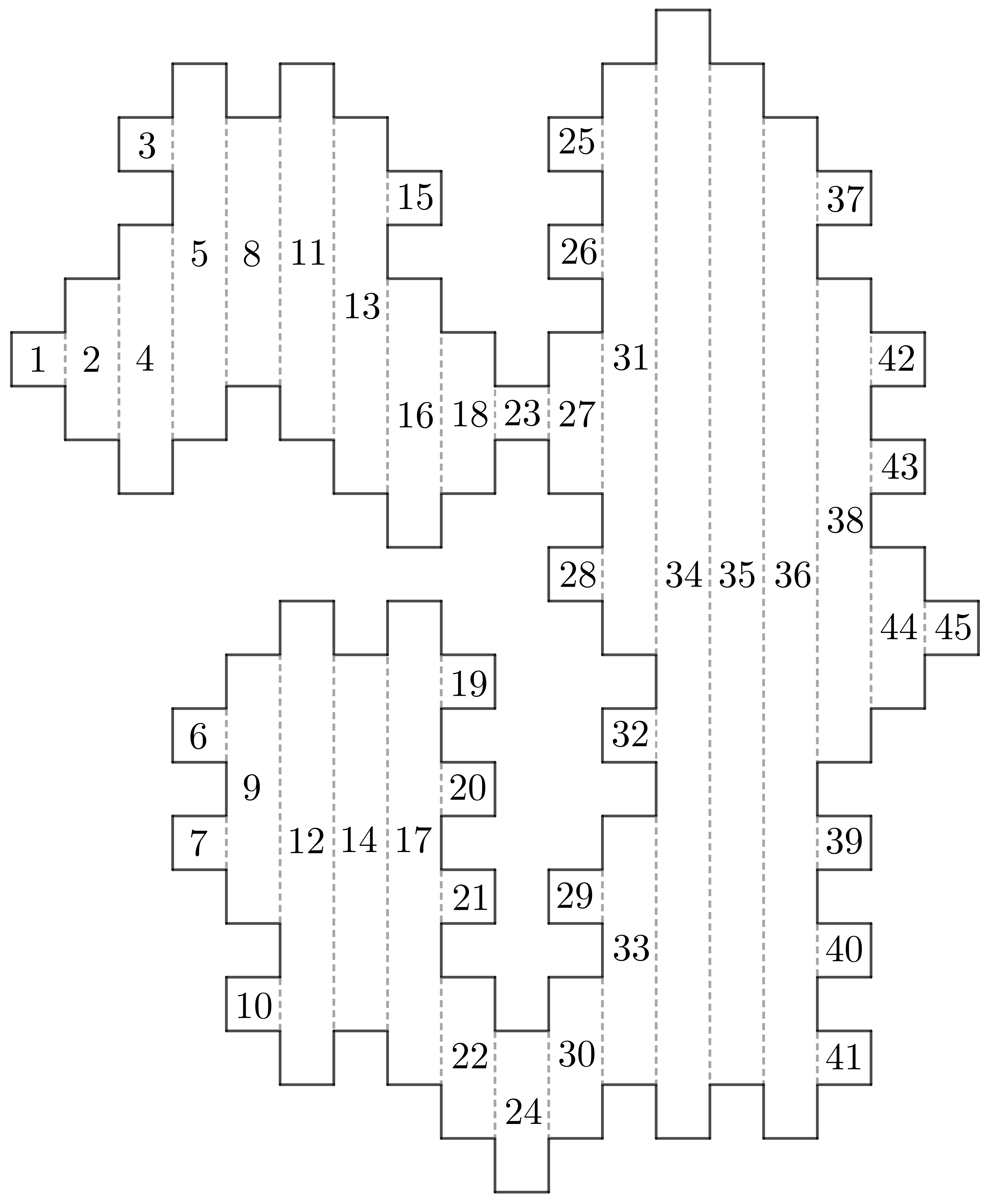}
\caption{Partitioning a polygon into columns. Columns like $1, 3, 37$ and $45$ are called teeth.}
\label{fig:column_tree_a}
\end{subfigure}
~
\begin{subfigure}{0.65\textwidth}
\centering
\includegraphics[width=\linewidth]{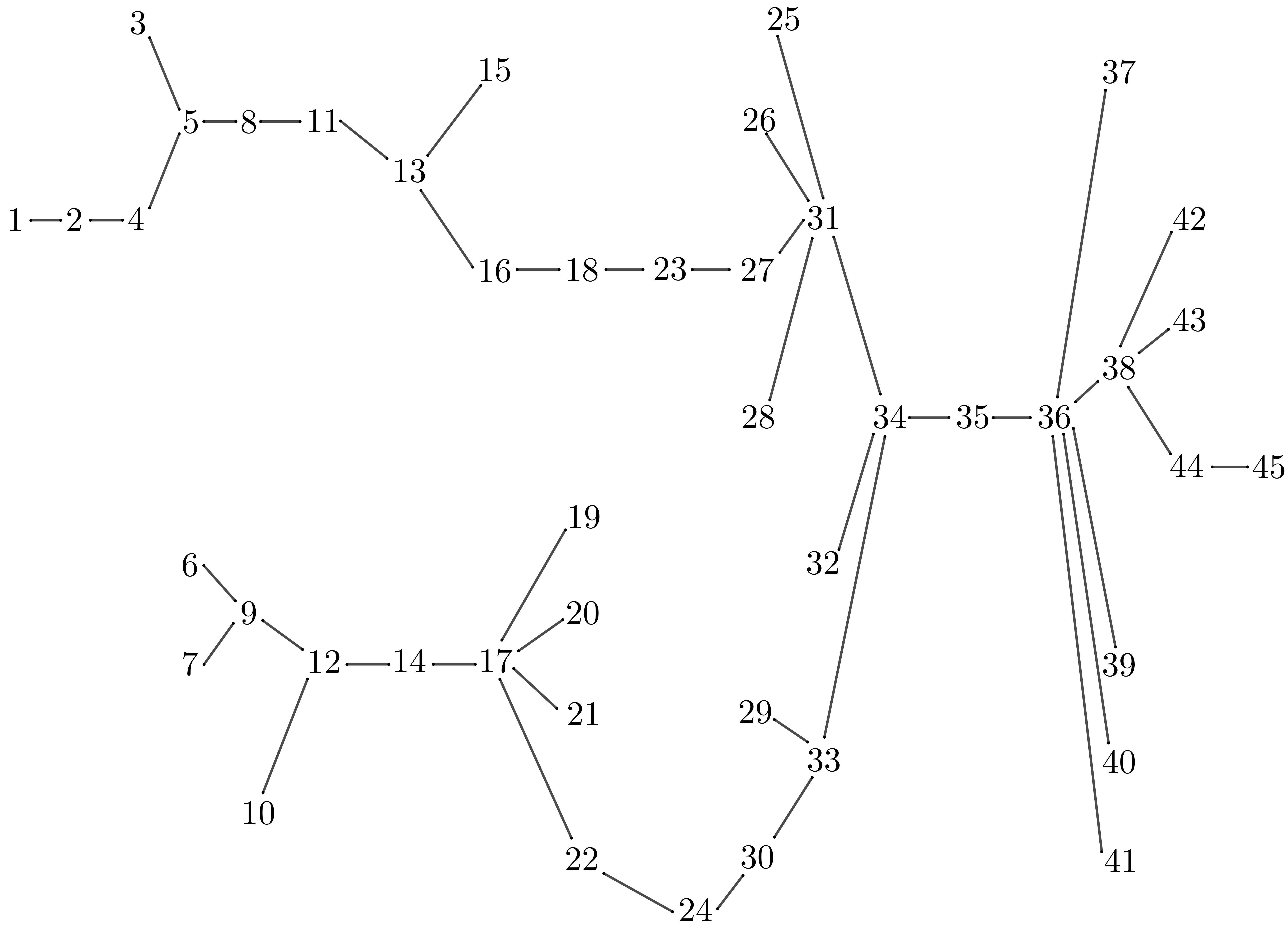}
\caption{The corresponding column tree. Nodes representing teeth are leaves in the column tree.}
\label{fig:column_tree_b}
\end{subfigure}

\caption{When two columns share a border in figure (a), their corresponding nodes share an edge in figure (b).}
\label{fig:column_tree}
\end{figure}

The \emph{column tree} of $P$, denoted by $T$, is the tree whose nodes are the $\frac{n}{4}$ columns of $P$, where two nodes of $T$ are joined by an edge if they are separated by a vertical cut.  
For an example see Figure~\ref{fig:column_tree}. 

%%%% Figure of column tree begin %%%%

%%%% Figure of column tree end %%%%

Because $P$ is ortho-unit, leaves in $T$ correspond to columns in $P$ that are one unit high, and have one edge of $P$ to their left or right side.  We call such structures \emph{teeth} in $P$; teeth can be either left- or right-facing depending on whether they have an edge of $P$ to their right or left side, respectively.  
As a second auxiliary structure, we create the \emph{tooth graph} $H$ of $P$ as follows: the nodes of $H$ are the teeth of $P$.  Two teeth $t_1$ and $t_2$ are connected by an edge in $H$ if the horizontal cut from either the top or bottom edge of $t_1$ coincides with a horizontal cut from the top or bottom edge of $t_2$.  Equivalently, two teeth are connected by an edge if a single guard can simultaneously guard both of them.  

An \emph{interior edge} of $T$ is an edge of $T$ which is not incident to a leaf.  
Each interior edge corresponds to a vertical cut which is \emph{not} incident to a tooth.  
We say that an interior edge of $T$ can guard a tooth if a guard can be placed on the corresponding vertical cut in such a way that it guards the tooth.

\section{Guarding ortho-unit polygons}

In this section we prove our main result, namely:

\begin{theorem} 
\label{thm:main}
Any ortho-unit polygon with $n \geq 12$ vertices can be guarded with at most $\lfloor \frac{n-4}{8} \rfloor$ guards.  
Moreover, $\lfloor \frac{n-4}{8} \rfloor$ guards are sometimes necessary.  
\end{theorem} 

The lower bound stated in Theorem~\ref{thm:main} is attained by the family of ortho-unit polygons shown in Figure~\ref{fig:macahuitl}.

\begin{figure}[ht!]
%\captionsetup{width=.9\linewidth}
\centering

\begin{subfigure}[b]{0.3\textwidth}
\centering
\begin{tikzpicture}[scale=.45, every node/.style={draw=none}]
\draw[] (0,1)--(0,0)--(1,0)--(1,-1)--(2,-1)--(2,0)--(3,0)--(3,-1)--(4,-1)--(4,0)--(5,0)--(5,-1)--(6,-1)--(6,0)--(7,0)

(0,1)--(1,1)--(1,2)--(2,2)--(2,1)--(3,1)--(3,2)--(4,2)--(4,1)--(5,1)--(5,2)--(6,2)--(6,1)--(7,1)--(7,0)
;

\end{tikzpicture}
\caption{}
\end{subfigure}
~
\begin{subfigure}[b]{0.4\textwidth}
\centering

\begin{tikzpicture}[scale=.45, every node/.style={draw=none}]
\draw[] (0,1)--(0,0)--(1,0)--(1,-1)--(2,-1)--(2,0)--(3,0)--(3,-1)--(4,-1)--(4,0)--(5,0)--(5,-1)--(6,-1)--(6,0)--(7,0)

(0,1)--(1,1)--(1,2)--(2,2)--(2,1)--(3,1)--(3,2)--(4,2)--(4,1)--(5,1)--(5,2)--(6,2)--(6,1)--(7,1)
;
\draw[gray, thick](7,1)--(7,2)--(8,2)--(8,1)--(9,1) (9,0)--(8,0)--(8,-1)--(7,-1)--(7,0);
\draw[] (9,1)--(9,0);

\end{tikzpicture}
\caption{}
\end{subfigure}

\caption{(a) is a short `macuahuitl' polygon, with 28 sides, requiring 3 guards. (b) is an illustration of extending a `macuahuitl' polygon. This requires 8 additional sides and one additional guard. Both polygons require $\frac{n-4}{8}$ guards.}
\label{fig:macahuitl}
\end{figure}

With respect to  the upper bound, note that a guard placed on a vertical cut corresponding to an edge of $T$ will guard both incident columns.  
A simple observation, then, is that a perfect matching in $T$ would yield a set of $\frac{n}{8}$ guards which guard all of $P$: each of the edges in the matching corresponds to a vertical cut, and placing a guard on each of these vertical cuts would result in a set of guards which guard the entire polygon.  
Of course, $T$ will typically not contain a perfect matching but this observation is our starting point, nonetheless.

The following proposition, which follows easily by induction, reveals that leaves of $T$ are the barriers to finding a perfect matching.  
We state it in a form which will be convenient to us later in the proof.  

\begin{prop}
Suppose $T$ is a tree, and $x \in V(T)$ is a leaf.  Then there is a matching in $T$ which saturates $x$ and all interior nodes of $T$. \label{prop:match} 
\end{prop}

The rest of the proof consists of several steps and involves a collection of subgraphs and matchings applied on them.
We thus present an overview prior to dealing with the details of the proof.

\subsection{Upper bound - Overview}

We now give an outline of our strategy to guard $P$ at most $\lfloor \frac{n-4}{8} \rfloor$ vertex guards.
\begin{enumerate}[itemsep = 5pt]
    
    \item \textit{Handle the leaves of $T$:} First, we address the leaves of $T$ by proving that the graph $H$ consists of paths and isolated nodes (Lemma~\ref{lem:paths}). 
    We could thus construct a matching in $H$ where, for each odd component, only one endpoint remains unguarded (which is done at the end of the argument).

    \item \textit{Match odd components of $H$ with interior edges of $T$:} We match each odd component of $H$ to an interior edge of $T$. 
    This allows us to place a guard covering both columns associated to the interior edge and an endpoint of its paired odd component (Lemma~\ref{lem:tmatch}). 
    We refer to this matching pairing leaves in odd components as the \emph{$\mathcal{L}$-matching}.

    \item  \textit{Remove gaps from the $\mathcal{L}$-matching}: We then adjust the previous matching to avoid having any unmatched interior edge of $T$ in the path between an $\mathcal{L}$-matched interior edge and its paired tooth.
    This allows us to easily handle the remaining columns afterwards.

    \item \textit{Count covered columns in matched components:} With the adjusted matching in place, we count the number of columns guarded within each connected component of the subgraph of $T$ induced by the $\mathcal{L}$-matching, which we denote as $\mathcal{F}'$. 
    We prove that the $\lvert X \rvert$ guards placed for any connected component of $\mathcal{F}'$ guard $2\lvert X \rvert + 1$ columns of $P$ in total (Lemma~\ref{lem:count}).

    \item \textit{Extract the remaining forest $\mathcal{F}$:} By removing from $T$ the nodes incident to $\mathcal{L}$-matched interior edges and their paired teeth we obtain a forest $\mathcal{F}$ whose nodes correspond to unguarded columns.
    This forest $\mathcal{F}$ contains a subgraph $H' \subseteq H$ composed of even-length paths.

    \item \textit{Apply Proposition~\ref{prop:match} to $\mathcal{F}$:} We choose an appropriate leaf of $T$ as its root. 
    Then, for each connected component of forest $\mathcal{F}$, we apply Proposition~\ref{prop:match} in such a way that it saturates its node closest to the root of $T$.
    The resulting matching is called the $\mathcal{F}$-matching.
    Remaining unguarded columns are leaves of components in $\mathcal{F}$, which may or not be teeth.

    \item \textit{Relate unguarded non-teeth nodes to matched components:}
    We then associate each unguarded non-teeth node in $\mathcal{F}$ with a distinct connected component of $\mathcal{F}'$, thus achieving an average of two guarded columns per guard (this is guaranteed by Lemmas~\ref{lem:component} and~\ref{lem:blame}).

    \item \textit{Cover the remaining teeth:}
    For the remaining teeth, we fix a perfect matching on $H'$ and use it to adjust or add guards as needed until all columns are guarded.
    We call this matching the $\mathcal{T}$-matching.

    \item \textit{Refine the column counting:}
    Finally, we confirm that, in every case, we can cover at least one more column than twice the number of guards placed, thereby achieving the desired upper bound.

\end{enumerate}

\subsection{Upper bound - Details}

\begin{lemma} 
\label{lem:paths}
The components of $H$ consist of paths and isolated nodes.
\end{lemma} 

\begin{proof} 
The maximum degree of a tooth in $H$ is clearly two as there are only two horizontal cuts incident to it, one each coming from its top and bottom edge.  Since the teeth have height one, either \emph{both} the top and bottom horizontal cuts of two teeth coincide, or the top cut of one tooth coincides with the bottom cut of its neighbor.  This immediately rules out cycles, and so each component is a path.
\end{proof}

%An \emph{interior edge} of $T$ is an edge of $T$ which is not incident to a leaf.  Each interior edge corresponds to a vertical cut which is \emph{not} incident to a tooth.  We say that an interior edge of $T$ can guard a tooth if a guard can be placed on the corresponding vertical cut in such a way that it guards the tooth.

The following lemma is the key to efficiently guarding the teeth of the polygon.  
\begin{lemma} Suppose $P$ is an ortho-unit polygon.  
Let $O$ denote the set of odd components of the tooth graph $H$ of $P$.  Then there is a matching between $O$ and the interior edges of $T$ which saturates $O$ and so that each matched interior edge can guard a leaf in the corresponding matched component.  
\label{lem:tmatch}
\end{lemma}   

\begin{figure}[ht]
\centering
%\captionsetup{width=.9\linewidth}
\begin{tikzpicture}[scale=.45, every node/.style={draw=none}]

\draw[](1,1)--(1,2)--(2,2)--(2,3)--(1,3)--(1,4)--(0,4)--(0,5)--(1,5)--(1,6)--(0,6)--(0,7)--(1,7)--(1,8)--(0,8)--(0,9)--(-1,9);
\draw[](1,1)--(2,1)--(2,0)--(3,0)--(3,1)--(4,1)--(4,0)--(5,0)--(5,1)--(6,1)--(6,2)--(5,2)--(5,3)--(6,3)--(6,4)--(5,4)--(5,5)--(6,5)--(6,6)--(5,6)--(5,7)--(6,7)--(6,8)--(5,8)--(5,9);

\draw[] plot [smooth, tension=1] coordinates { (-1,9)
(-1.25,9.5)
(-.75,10)
(-1.25,10.5)
        };
\draw[] plot [smooth, tension=1] coordinates { (5,9)
(5.25,9.5)
(4.75,10)
(5.25,10.5)
        };

\node[blue] (a) at (.5,6.5) {$a$};
\node[red] (b) at (.5,4.5) {$b$};
\node[black!60!green] (c) at (1.5,1.5) {$c$};
\node[blue] (e) at (5.5,7.5) {$d$};
\node[red] (d) at (5.5,5.5) {$e$};
\node (f) at (5.5,3.5) {$f$};
\node[black!60!green] (i) at (5.5,1.5) {$g$};

\draw[dashed, black!60!green, very thick,] (2,2)--(5,2);

\draw[dashed, red, very thick,] (1,5)--(5,5);

\draw[dashed, blue, very thick,] (1,7)--(5,7);

\end{tikzpicture}
\caption{Above is a matching of teeth. We see that $c$ and $g$ are matched, as are $b$ and $e$ and, $a$ and $d$. $f$ is unmatched.}
\label{fig:teeth}
\end{figure}
\begin{proof} 
Components in $O$ are either isolated teeth or odd length paths of teeth.  Note that odd length paths can be arranged vertically, and there is a top-most and bottom-most edge of the top and bottom teeth of the path. 

We build a bipartite graph where one partite set consists of components in $O$ and the other partite set consists of interior edges of $T$ as follows.  Fix a component $x$ in $O$.  This component has a top-most and bottom-most horizontal edge, if the component consists of only one tooth, then these are the top and bottom edges of that tooth, otherwise they lie in the top and bottom tooth respectively.  
Consider the horizontal cut defined by the top-most edge of $x$; this intersects $P$ again at a reflex vertex, $r$.  
Because this is the topmost edge in a component of $H$, $r$ does not lie on a tooth.  
Hence, the vertical cut through $r$ corresponds to an internal edge $e$ of $T$.  
We connect the component $x$ to the internal edge $e$. 
In the same way, connect $x$ to an internal edge of $T$ using the bottom edge of $x$.
An example is shown in Figure~\ref{fig:matching}.

\begin{figure}[!htb]
\centering

\begin{subfigure}{0.65\textwidth}
\centering

\includegraphics[width=0.72\linewidth]{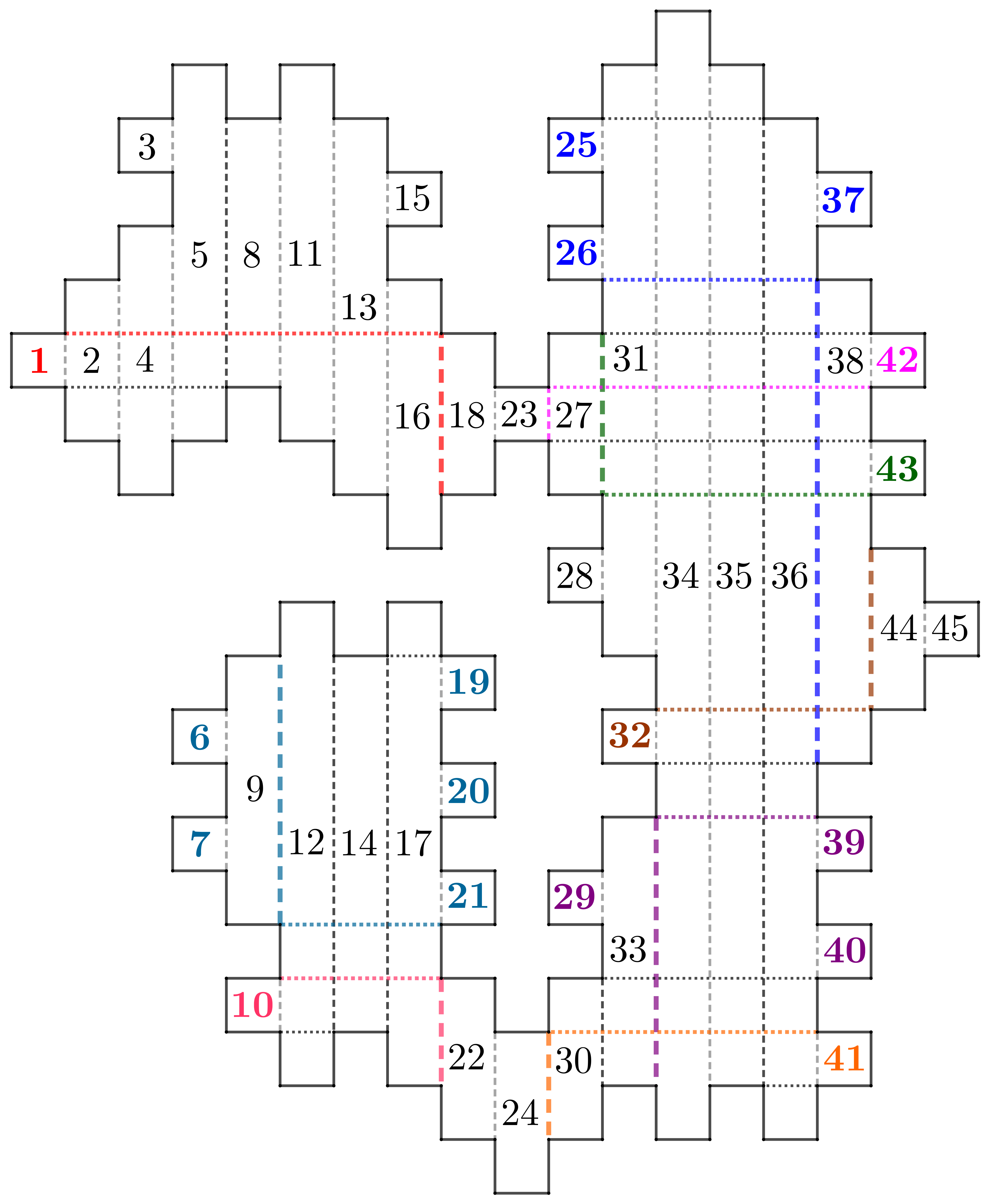}
\caption{A matching of internal edges and odd components of teeth using Hall's Theorem.}
%  \label{fig:}
\end{subfigure}
%\qquad
\begin{subfigure}{0.65\textwidth}
\centering
\includegraphics[width=0.52\linewidth]{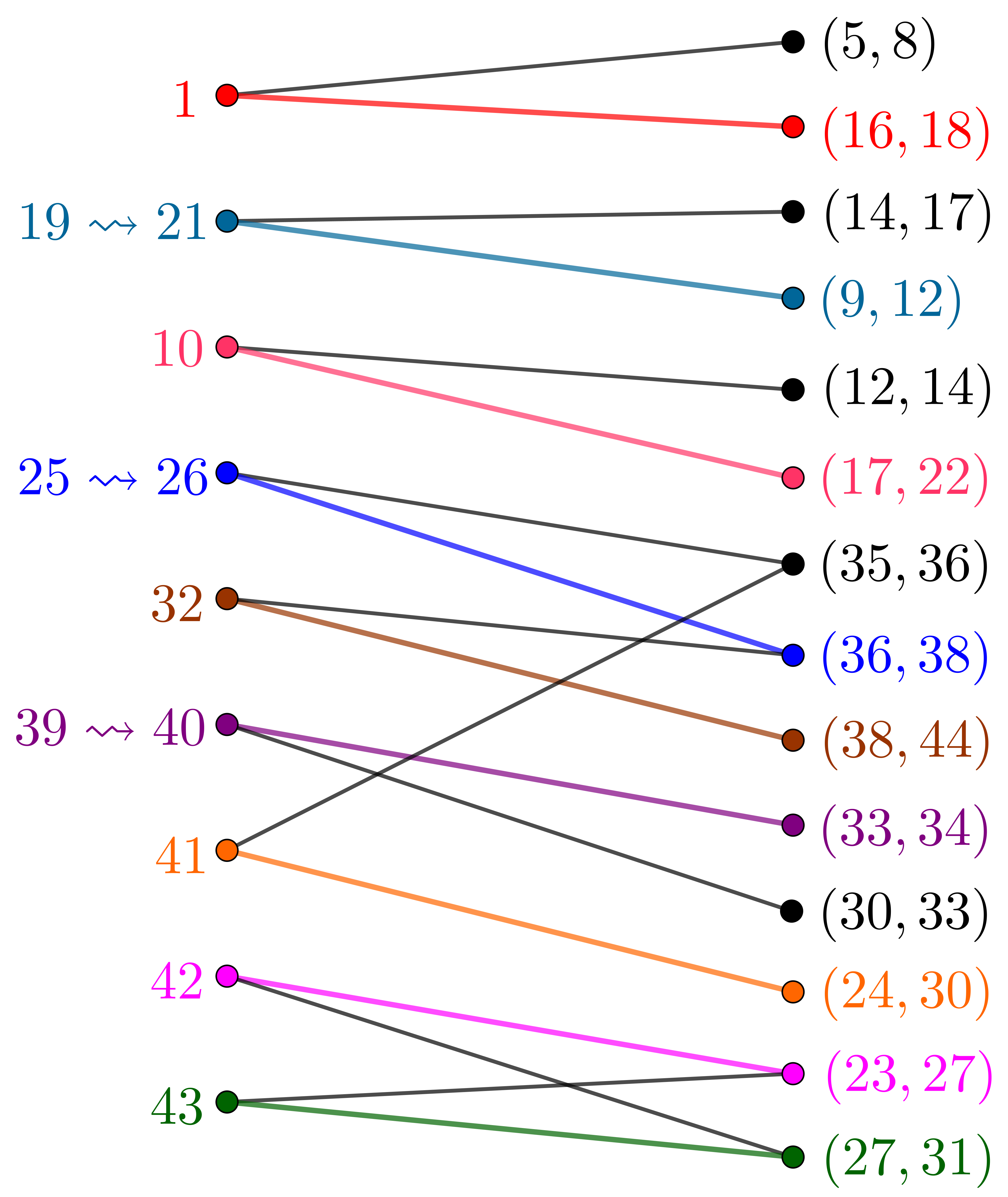}
\caption{A representation of the matching in the column tree. Black nodes correspond to unmatched internal edges.}
%\label{fig:}
\end{subfigure}
\caption{Illustration of the proof of Lemma~\ref{lem:tmatch}.}
\label{fig:matching}
\end{figure}

The internal edges have maximum degree two, their potential neighbors being determined by the horizontal cuts through the reflex vertices at the top and bottom.  Meanwhile, the odd components have degree \emph{exactly} two, unless the top and bottom of the component connect to the same interior edge (and hence the component is the only neighbor of the edge.)  
The existence of the desired matching, then, is an easy consequence of Hall's theorem \cite{hall1987representatives}.  

That the edge guards a tooth it is matched to is a consequence of the fact that one horizontal cut coming from the tooth intersects the vertical cut coming corresponding to such edge. 
%\textcolor{red}{We could say that the guard can be placed precisely at the reflex vertex we are referring to, thus obtaining a vertex guard as suggested by the reviewer.}
\end{proof}

We think of the matching guaranteed by a Lemma \ref{lem:tmatch} as a matching between leaves of $T$ (teeth) and internal edges so that one of the endpoints of each odd component of $H$ is matched.

\begin{figure}[!htb]
\centering

\begin{subfigure}[b]{0.65\textwidth}
\centering
\includegraphics[width=0.7\linewidth]{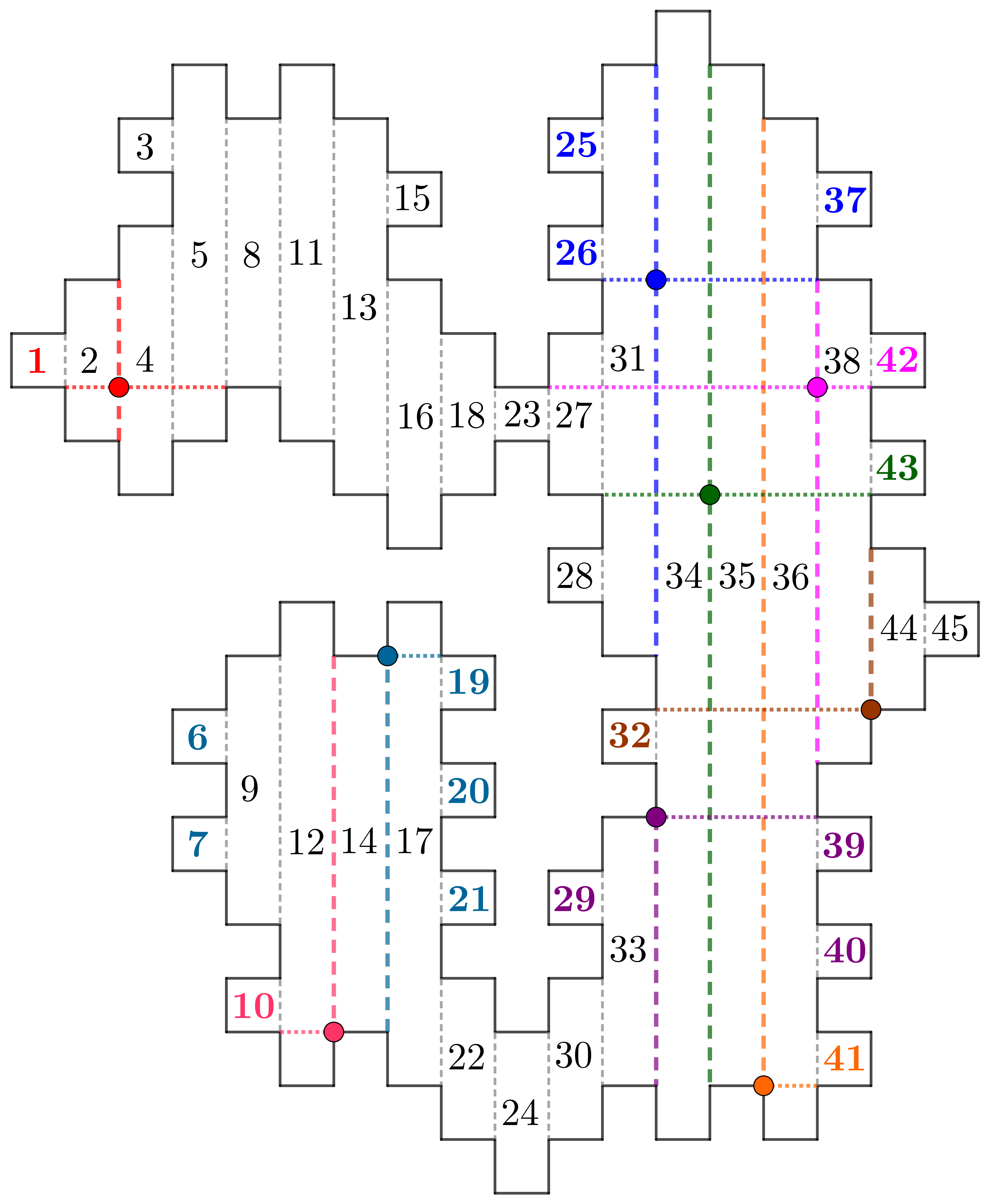}
\caption{A matching which minimizes the sum of distances from the teeth to the matched internal edges.}
%  \label{fig:}
\end{subfigure}
\qquad
\begin{subfigure}[b]{0.65\textwidth}
\centering

\includegraphics[width=\linewidth]{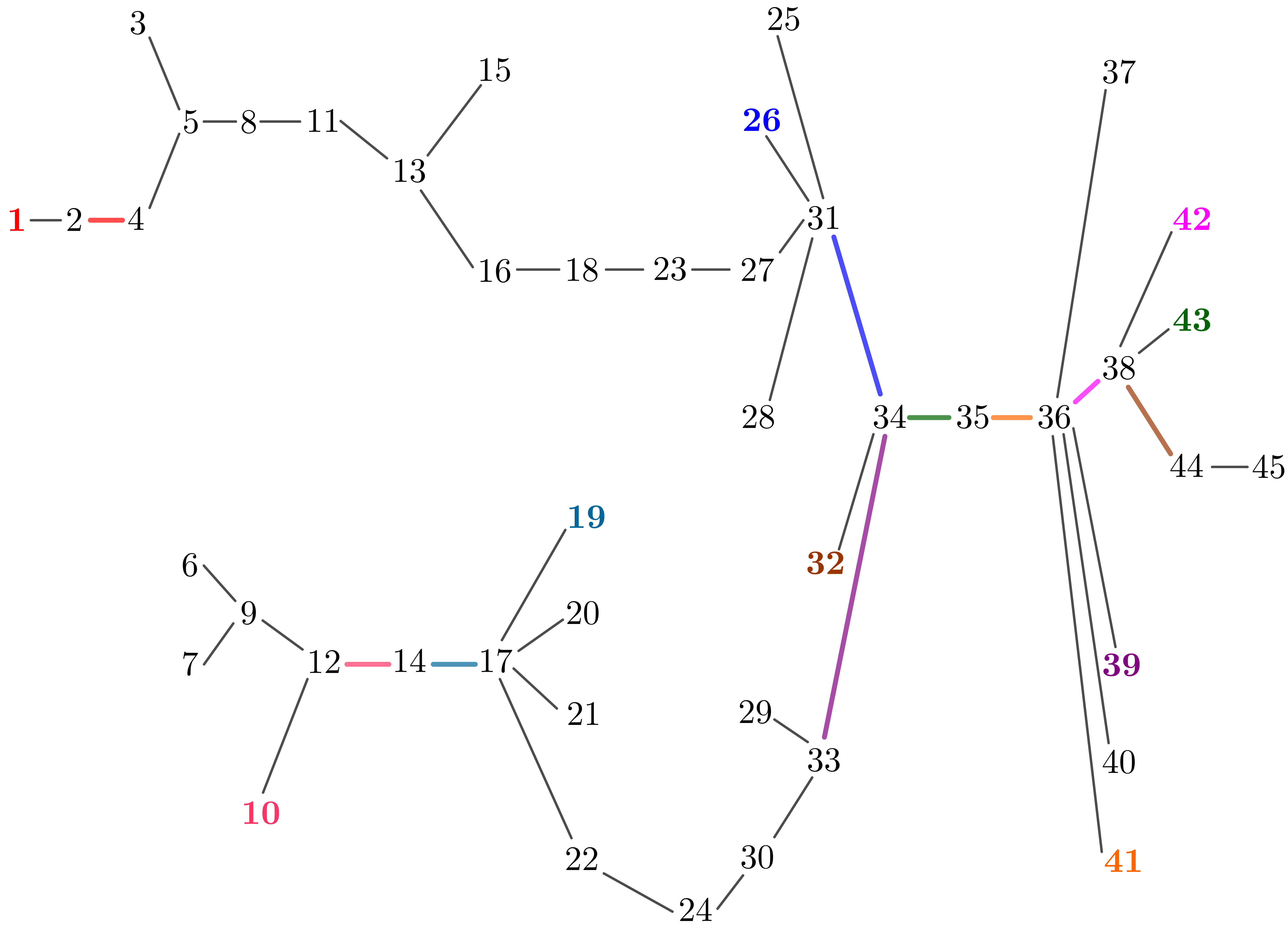}
%\caption{A representation of the matching in the column tree. Black nodes correspond to unmatched internal edges.}
\caption{$\mathcal{L}$-matched teeth and internal edges are shown bold and colored.}
%\label{fig:}
\end{subfigure}
\caption{A guard placement corresponding to a minimum-length $\mathcal{L}$-matching.}
\label{fig:minimum_matching}
\end{figure}

Now we are ready to turn to the meat of the proof of Theorem \ref{thm:main}.  For our ortho-unit polygon $P$, consider a matching guaranteed by Lemma \ref{lem:tmatch}.  %As several matchings are considered in the proof, we call this one the \textcolor{blue}{$\mathcal{L}$-matching}.  
While the matching the proof provides may match teeth with internal edges far from them, it is easier to deal with a `closer' matching.  
%To that end, we consider a matching satisfying the conclusion of the lemma that minimizes the sum of distances from the teeth (leaves) to the matched internal edges, see Figure~\ref{fig:minimum_matching}.  
To that end, we first augment the bipartite graph from the proof of Lemma~\ref{lem:tmatch} by connecting each odd component in $H$ to all internal edges intersected by its top and bottom horizontal cut.
We then consider a matching satisfying the conclusion of Lemma~\ref{lem:tmatch} that minimizes the sum of distances from the teeth (leaves) to the matched internal edges, see Figure~\ref{fig:minimum_matching}.
Such a matching has the property that any internal edge in the path from a tooth to its matched internal edge is in turn matched to a tooth (as otherwise there is a matching whose sum of distances is smaller).
%\textcolor{red}{This implies that no endpoint of a unmatched internal edge sits next to a matched tooth.}
Fix such a minimal matching; we will work with it for the remainder of the proof.  
As several matchings are considered in the proof, we call this one the $\mathcal{L}$-matching.

%Colored Longer Column Tree Figure

\begin{figure}[ht]
\centering
\includegraphics[width=0.65\linewidth]{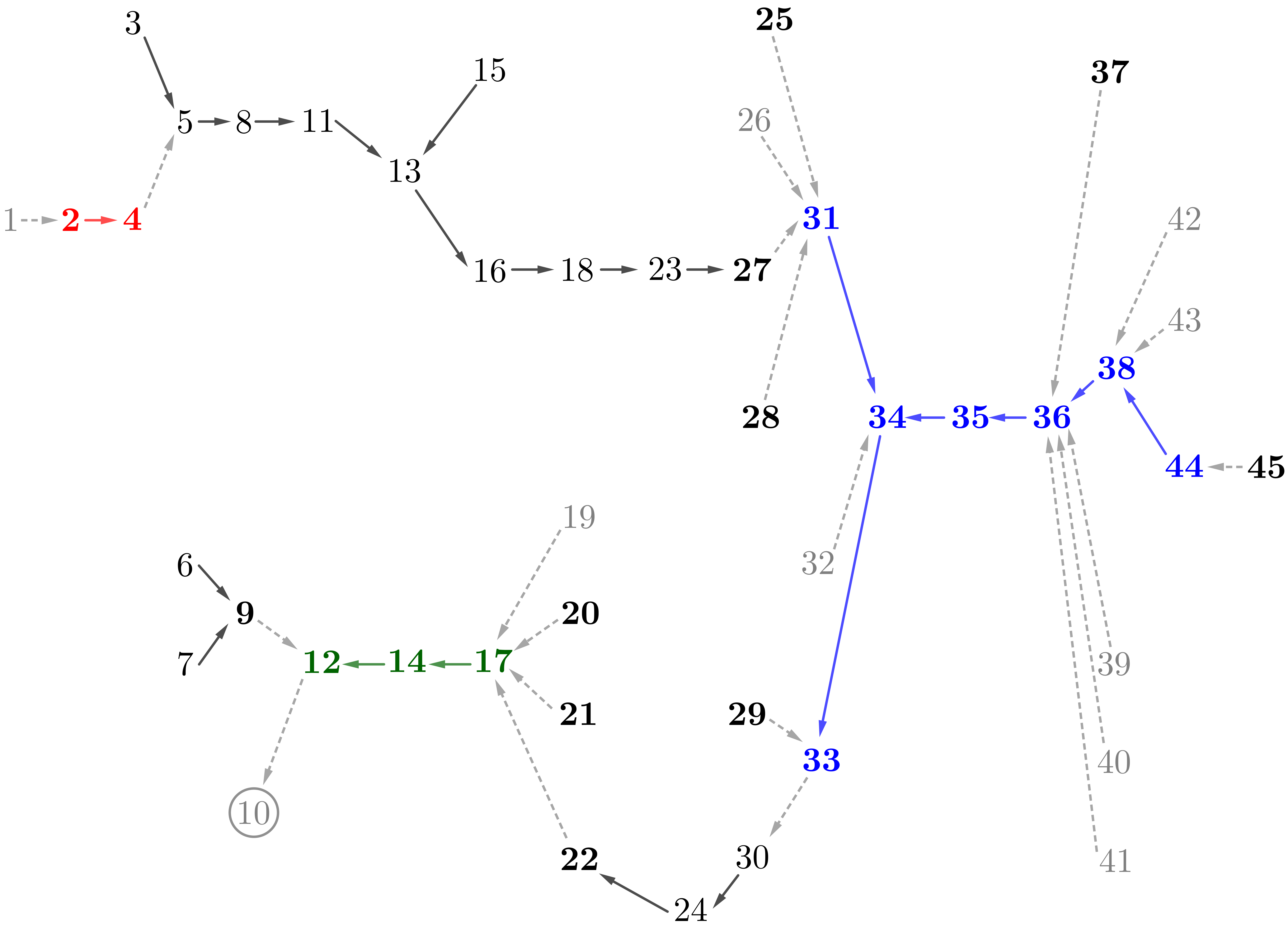}
\caption{Suppose the column tree is rooted at $10$. The red, blue, and green subtrees are the components of $\mathcal{F}'$. Black nodes are in $\mathcal{F}$, the bold ones being the roots of their corresponding components. Grey nodes are $\mathcal{L}$-matched teeth, which are not included in $\mathcal{F}$ or $\mathcal{F}'$. While $30$ blames the blue component, $5$ does not blame the red component as it is not a leaf of its component of $\mathcal{F}$. The green component remains unblamed.}
\label{fig:forest}
\end{figure}

Place guards on the vertical cuts corresponding to $\mathcal{L}$-matched internal edges in such a way that they also guard their matched tooth.  
We note that, because of the minimality condition of the $\mathcal{L}$-matching, these cannot necessarily be placed at vertices.
Let $\mathcal{F}'$ be the forest spanned by $\mathcal{L}$-matched internal edges in $T$, and $\mathcal{F}$ be the forest obtained by removing all nodes in $\mathcal{F}'$ along with the $\mathcal{L}$-matched teeth, see Figure~\ref{fig:forest} for an example.  
%\textcolor{blue}{Let $\mathcal{F}'$ be the forest in $T$ induced by nodes of the $\mathcal{L}$-matched internal edges and the $\mathcal{L}$-matched teeth, and $\mathcal{F}$ be the forest obtained by removing all nodes in $\mathcal{F}'$ from $T$, see Figure~\ref{fig:forest}.} 
The nodes in $\mathcal{F}$, then, correspond to columns that still need to be guarded.  Denote by $H'$ the graph obtained by removing all $\mathcal{L}$-matched teeth from the tooth graph $H$.  Note that $H'$ consists solely of even paths, and hence contains a perfect matching, which we call the $\mathcal{T}$-matching.

%As we consider a minimal-distance $\mathcal{L}$-matching, $\mathcal{L}$-matched teeth are contained in the same connected component of $\mathcal{F}'$ as the endpoints of their paired internal edges.
%Hence, each component of $\mathcal{F}'$ contains as many {$\mathcal{L}$-matched} internal edges as teeth.
%Now we count the number of columns guarded in each such component of $\mathcal{F}'$.

\begin{lemma}
	Let $X$ be the set of $\mathcal{L}$-matched internal edges spanning some component of $\mathcal{F'}$.  
    Then the guards placed on the corresponding vertical cuts guard a total of $2|X|+1$ columns of $P$, including their paired teeth.  
    \label{lem:count} 
\end{lemma}   
\begin{proof}
Each guard sees both columns incident to the vertical cut it sits on.  Since the internal edges span a tree whose nodes are these columns, this is precisely $|X|+1$ columns.  
Furthermore, each guard also guards the tooth $\mathcal{L}$-matched to its internal edge.  Since all edges are internal, these teeth are disjoint from the columns already counted and $2|X|+1$ columns are guarded in total. 
%Since all edges are internal, these are disjoint from the columns already counted and $2|X|+1$ columns are guarded in total.  
\end{proof}

Now we need to guard the remaining columns corresponding to nodes in $\mathcal{F}$.  To this end, we fix a root in $T$.  If $\mathcal{F}'$ is non-empty, we choose the root to be a leaf $\mathcal{L}$-matched to an internal edge in $\mathcal{F}'$.  If all components in $H$ are even, then $\mathcal{F}'$ is empty; in this case, we choose the root to be an arbitrary leaf.  We now define, for every component of $\mathcal{F}$, its root to be the node in the component closest to the root of $T$. 

Let $\ell$ be a leaf of a component of $\mathcal{F}$. 
%We say $\ell$ \emph{blames} a component $C$ of $\mathcal{F}'$ if $\ell$ is the first node outside of $C$ on the shortest path from $C$ to the root in $T$.  
%We say $\ell$ \emph{blames} a component $C$ of $\mathcal{F}'$ if $\ell$ is the first node \textcolor{blue}{of $\mathcal{F}$} on the path from $C$ to the root of $T$.  
We say $\ell$ \emph{blames} a component $C$ of $\mathcal{F}'$ if $\ell$ is adjacent to $C$ and lies on the path of $T$ between $C$ and the root of $T$.
From the definition, it immediately follows
\begin{lemma}
No component $C$ of $\mathcal{F}'$ is blamed by more than one leaf $\ell$.  
\label{lem:component}
\end{lemma}

%Blaming Figure

On the other hand,
\begin{lemma}
Suppose $\ell$ is a leaf in $\mathcal{F}$ that is not a leaf in $T$.  If $\ell$ is isolated or not the root of its component, then $\ell$ blames some component $C$ in $\mathcal{F'}$.  
\label{lem:blame}
\end{lemma} 

\begin{proof} 
Because $\ell$ is not a leaf in $T$, it is an internal node. 
Because $T$ is a tree, $\ell$ lies on the shortest path from a leaf to the root of $T$.  
Because $\ell$ is either isolated or not the root of its component, the node of distance one from $\ell$ to this leaf is not in $\mathcal{F}$. %, otherwise $\ell$ would not be a leaf in $\mathcal{F}$. 
%The node of distance one from $\ell$ to this leaf is not in $\mathcal{F}$, else $\ell$ would not be a leaf, or not be isolated in $\mathcal{F}$. 
Moreover, as the $\mathcal{L}$-matching minimizes the sum of distances between $\mathcal{L}$-matched leaves and internal edges, this node is not an $\mathcal{L}$-matched leaf of $T$.
Therefore, this node belongs to some component $C$ of $\mathcal{F'}$. 
Since $\ell$ is distance one from $C$ and lies on the shortest path from $C$ to the root of $T$, $\ell$ blames $C$.
\end{proof}

We are now ready to complete the proof.

\begin{proof}[Proof of Theorem \ref{thm:main}]

It remains to guard the nodes of $\mathcal{F}$. For each non-trivial component, we apply Proposition \ref{prop:match} to obtain a matching in $\mathcal{F}$, the $\mathcal{F}$-matching, which covers all interior nodes of the component and the root of each component (taking that root to be $x$ in Proposition \ref{prop:match} if the root is a leaf).  We place guards on each of the vertical cuts corresponding to these matched edges.  At this point, we have placed guards on the vertical cuts of the edges of $\mathcal{F'}$ and the $\mathcal{F}$-matched edges of $\mathcal{F}$.  The only unguarded columns now correspond to isolated nodes and unmatched leaves in $\mathcal{F}$.  

%Forest matching
\begin{figure}[!htb]
\centering

\begin{subfigure}[b]{0.65\textwidth}
\centering
\includegraphics[width=0.7\linewidth]{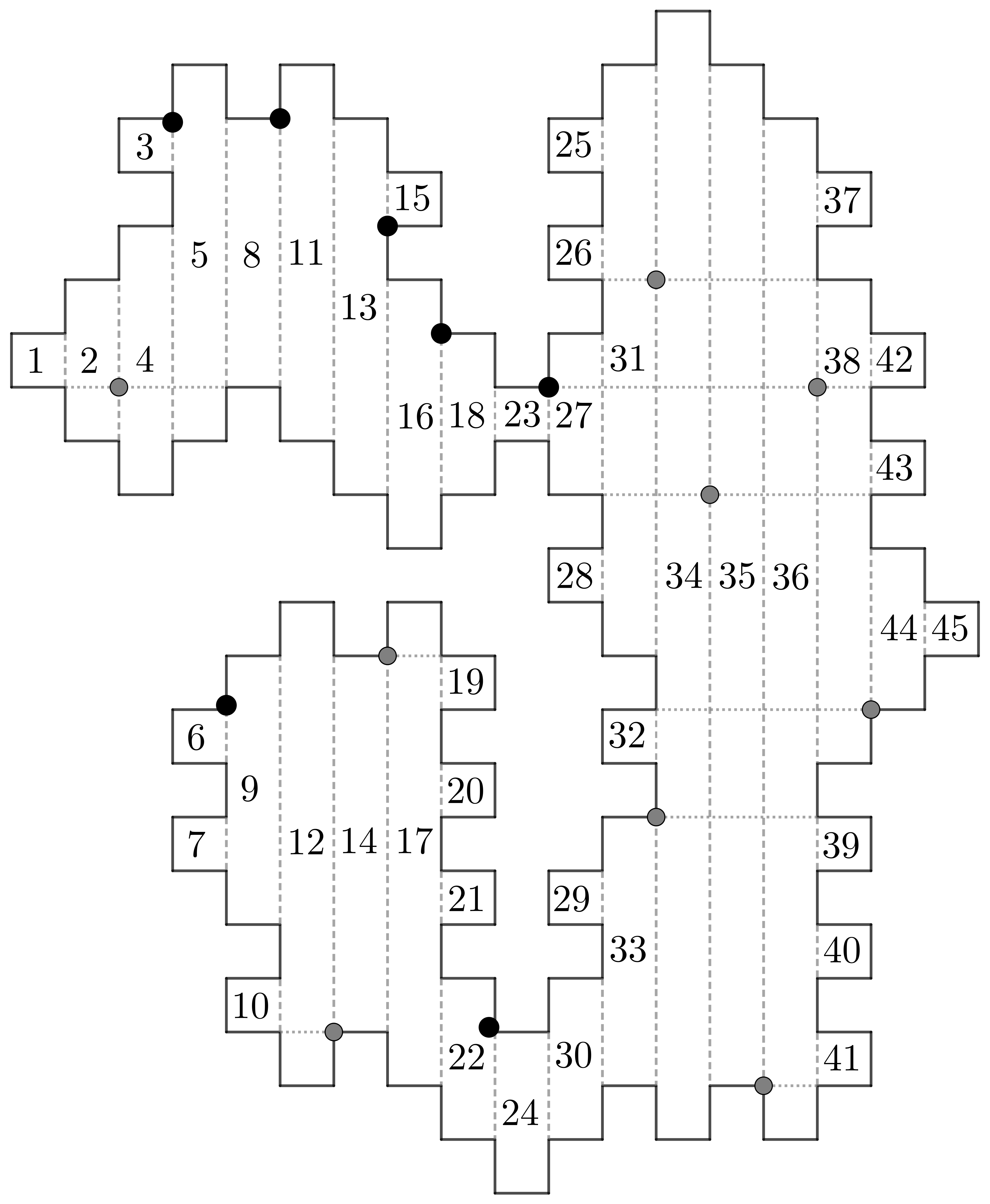}
\caption{Black dots represent a guard placing corresponding to the $\mathcal{F}$-matched pairs.}
%  \label{fig:}
\end{subfigure}
\qquad
\begin{subfigure}[b]{0.65\textwidth}
\centering
\includegraphics[width=\linewidth]{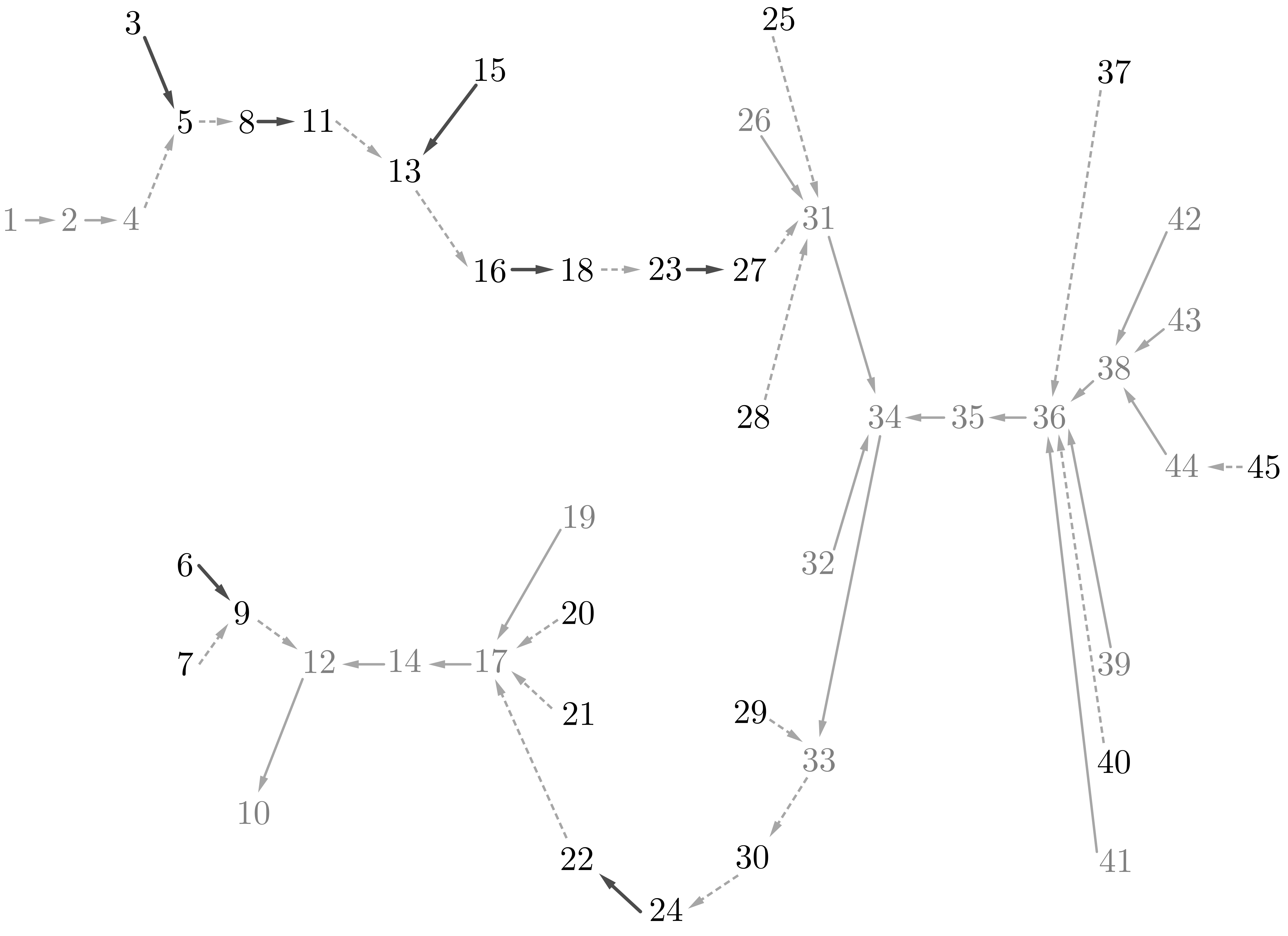}
\caption{The nodes of $\mathcal{F}$ are labeled black,  $\mathcal{F}$-matched nodes are joined with a black solid arrow.}
%\label{fig:}
\end{subfigure}
\caption{$\mathcal{F}$-matching in the components of $\mathcal{F}$. }
\label{fig:forest_matching}
\end{figure}

For each unmatched leaf or isolated node in $\mathcal{F}$ that is not a leaf in $T$ we place a guard dedicated to guarding the associated column.  

The remaining leaves correspond to teeth in $P$ and are handled as follows: fix a perfect $\mathcal{T}$-matching in $H'$, the tooth graph consisting of teeth that were not matched in the first step.  For each edge of this $\mathcal{T}$-matching, we add guards as follows:
\begin{enumerate}[label=(\alph*)]
\item If both teeth are saturated in the $\mathcal{F}$-matching we do nothing, as both teeth are already guarded.  
\item If one tooth is saturated in the $\mathcal{F}$-matching but one is not, note that the guard can be placed in such a way that it actually guards \emph{both} teeth simultaneously, while still guarding the columns it was originally responsible for. %\textcolor{red}{The saturated tooth is matched to an adjacent column which can be guarded by any of the two reflex vertices of the tooth; choose the one in the horizontal cut shared with the unsaturated tooth.}
\item If neither tooth is saturated in the $\mathcal{F}$-matching, place a single guard responsible for both teeth.  %\textcolor{red}{The guard is placed at a vertex in their common horizontal cut.}
\end{enumerate} 

\begin{figure}[!htb]
\centering

\includegraphics[width=0.55\linewidth]{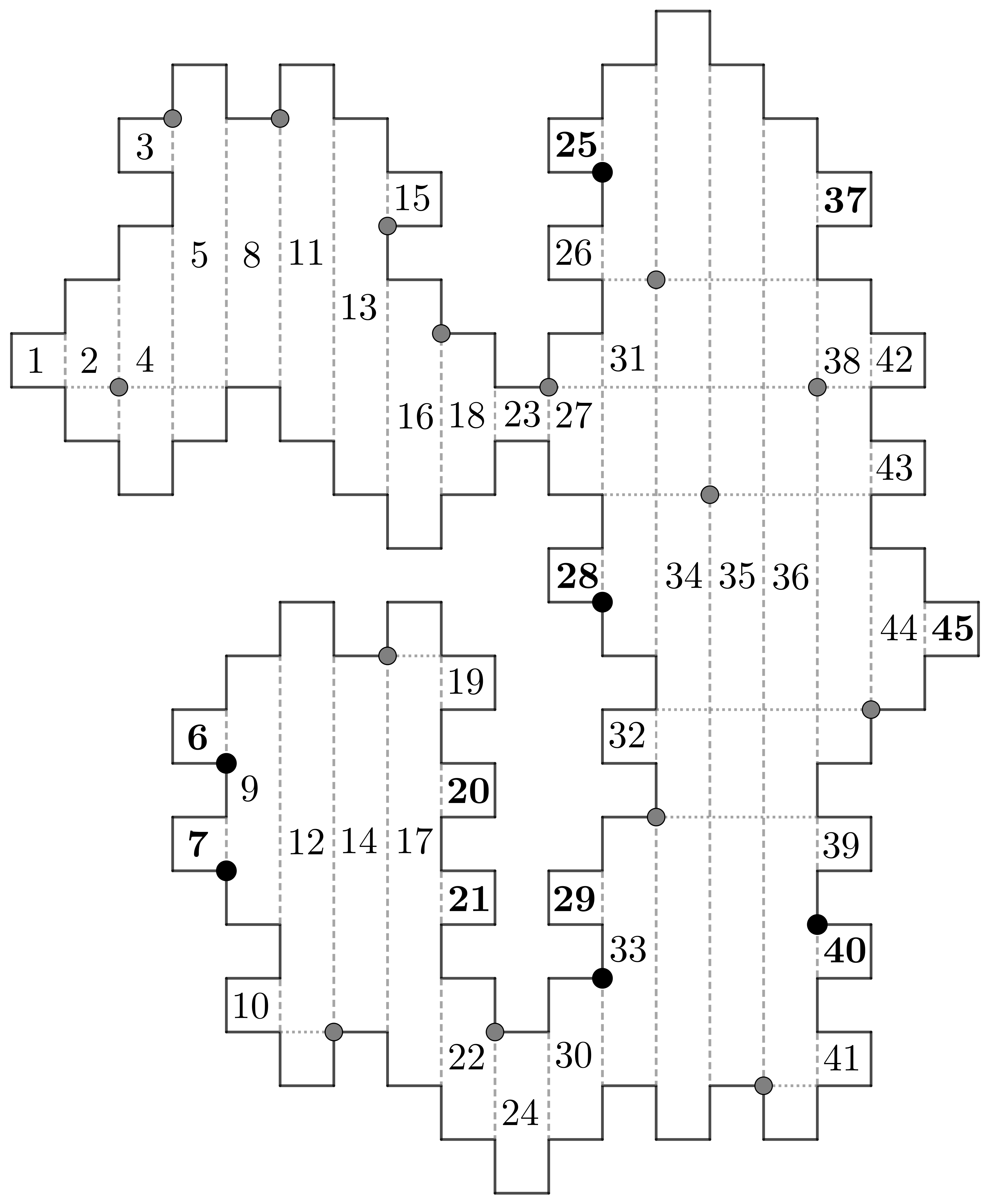}

\caption{Isolated node 30 is assigned a guard which also covers 33. The rest of the black dots are guards placed or adjusted by the $\mathcal{T}$-matching. 
Teeth in even-length paths are shown with bold labels.}
\label{fig:even_matching}
\end{figure}

By this construction all columns are guarded.  Furthermore, on average each guard is responsible for at least two unique columns.  This follows as the only guards initially assigned a single column are those which, by Lemma \ref{lem:blame} blame some component of $\mathcal{F}'$.  Per Lemma \ref{lem:count}, the guards guarding the columns of a component of $\mathcal{F}'$ guard, in the aggregate, an extra column beyond the claimed average of two each.   
Assigning this `extra' column to the guard only assigned a single column which blames that component gives the claimed average.  

This suffices to prove that at most $\lfloor \frac{n}{8} \rfloor$ guards can guard $P$. To prove that, actually, at most $\lfloor \frac{n-4}{8} \rfloor$ guards suffice, we proceed as follows.  

If $\mathcal{F}'$ is non-empty, note that the root of $T$ is incident to an internal node $x$ of $T$ that by our choice of matching is in $\mathcal{F}'$.  The component of $x$ is not blamed by any leaf and hence the `extra' column is never assigned.  Thus, if $g$ guards are placed $2g + 1$ is a lower bound for the number of distinct columns guarded so that $2g+1 \leq \lfloor \frac{n}{4} \rfloor$; rearranging $g \leq \lfloor \frac{n-4}{8} \rfloor$ as claimed.  

If $\mathcal{F}'$ is empty, which occurs if all components of $H$ are even, there is an `extra' column assigned to a guard if, when considering the $\mathcal{T}$-matching, case (b) above ever occurs. 

If case (b) never occurs, take an alternating path in the matching on $\mathcal{F}$ starting from some leaf $\ell$ in $T$.  This ends at some other leaf of $T$, $\ell'$. 

If $\ell$ and $\ell'$ are not paired in the $\mathcal{T}$-matching with each other, interchange the edges of the alternating path to obtain a new matching.  For this matching it is easy to see that case (b) will occur and the matching satisfies all other desired conditions.

If $\ell$ and $\ell'$ are paired in the $\mathcal{T}$-matching, ensure that we are in case (c) above, possibly interchanging the edges of the alternating path to obtain a new $\mathcal{F}$-matching.  Note that the path between $\ell$ and $\ell'$ in $T$ then contains a matched internal edge, and a guard can be placed on this internal edge in such a way that it sees not only the incident columns but  also guards \emph{both} the teeth corresponding to both $\ell$ and $\ell'$.  

This assignment of guards to columns ensures that there is at least one more than twice the number of guards columns guarded, and hence yields the desired bound.  

\end{proof}

%%%%%%%%%%%%%%%%%%%%%%%%%%%%%%%%%%

%\begin{comment}

\section{Guarding integral polygons}

An orthogonal polygon is an \emph{integral orthogonal polygon} if all of its edges have integer length. 
The lower bound on the number of guards needed to guard this family of polygons is given by the minimum-perimeter integral version of the well-known orthogonal comb polygon with $n$ vertices which requires $n/4$ guards.
This polygon has perimeter $N$ and requires $\lfloor N/6 \rfloor$ guards.
See Figure~\ref{fig:integral_lower_bound} for an example.

\begin{figure}[ht]
\centering
\begin{tikzpicture}[scale=.73, every node/.style={draw=none}]

\draw[](5,0)--(0,0)--(0,2)--(1,2)--(1,1)--(2,1)--(2,2)--(03,2)--(3,1)--(4,1)--(4,2)--(5,2)--(5,1);

\draw[](6,1)--(6,2)--(7,2)--(7,0)--(6,0);

\draw[dashed](5,1)--(6,1);
\draw[dashed](5,0)--(6,0);

\draw[fill=black] (0,0) circle (2pt);
\draw[fill=black] (1,0) circle (2pt);
\draw[fill=black] (2,0) circle (2pt);
\draw[fill=black] (3,0) circle (2pt);
\draw[fill=black] (4,0) circle (2pt);
\draw[fill=black] (5,0) circle (2pt);
\draw[fill=black] (6,0) circle (2pt);
\draw[fill=black] (7,0) circle (2pt);
\draw[fill=black] (0,1) circle (2pt);
\draw[fill=black] (1,1) circle (2pt);
\draw[fill=black] (2,1) circle (2pt);
\draw[fill=black] (3,1) circle (2pt);
\draw[fill=black] (4,1) circle (2pt);
\draw[fill=black] (5,1) circle (2pt);
\draw[fill=black] (6,1) circle (2pt);
\draw[fill=black] (7,1) circle (2pt);
\draw[fill=black] (0,2) circle (2pt);
\draw[fill=black] (1,2) circle (2pt);
\draw[fill=black] (2,2) circle (2pt);
\draw[fill=black] (3,2) circle (2pt);
\draw[fill=black] (4,2) circle (2pt);
\draw[fill=black] (5,2) circle (2pt);
\draw[fill=black] (6,2) circle (2pt);
\draw[fill=black] (7,2) circle (2pt);

\end{tikzpicture}

\caption{An integral polygon $P$ of perimeter $N$ which requires $\lfloor \frac{N}{6} \rfloor$ guards. Dots represent points with integer coordinates.
}
\label{fig:integral_lower_bound}
\end{figure}

We can prove, via an unenlightening proof by cases (not included in this paper) that integral orthogonal polygons of perimeter $N$ can be guarded using at most $\lfloor \frac{N}{5} \rfloor$ guards. 
%We conjecture, however, that $\lfloor \frac{N}{6} \rfloor$ guards are always sufficient:

\begin{conj*}
Let $P$ be an integral orthogonal polygon of perimeter $N$. Then $P$ can always be guarded with at most $\lfloor \frac{N}{6} \rfloor$ guards.
\end{conj*}

\bibliography{refs}
\bibliographystyle{abbrv}

\end{document}